\documentclass[epj,final]{svjour}
\usepackage{amsfonts}
\usepackage{amsmath}
\usepackage{amssymb}
\usepackage{graphics}
\begin{document}
\title{Robust Dynamical Recurrences Based on Floquet Spectrum}
\author{Muhammad Ayub\inst{1,2} \and Khalid Naseer\inst{1,3}  \and Farhan Saif\inst{1}% etc
}                     % Do not remove
\mail{ayubok@yahoo.com}
\institute{Department of Electronics, Quaid-i-Azam University, 45320, Islamabad, Pakistan.\and Theoretical Plasma Physics Division, PINSTECH, Nilore, Islamabad, Pakistan.\and Department of Physics, University of Sargodha, Sargodha, Pakistan.}
\date{Received: date / Revised version: date}
% The correct dates will be entered by Springer
%
\abstract{
Recurrence behavior of wave packets in coupled higher dimensional
systems and periodically driven systems is analyzed, which takes place in the
realm of higher coupling/modulation strength. We analyze the wave packet
dynamics close to nonlinear resonances developed in the systems and provide the
analytical understanding of recurrence times. We apply these analytical results
to investigate the recurrence times of matter
waves in optical lattice in the presence of external periodic forcing. The
obtained analytical results can experimentally be observed using currently 
available experimental setups.
} %end of abstract
\maketitle
\section{Introduction}In one dimensional quantum systems with discrete energy
spectrum, quantum recurrence theorem establishes the existence of recurrent
dynamics \cite{PR1957}. Quantum characteristics of dynamical systems, which
exhibit chaos in their
classical domain, have posed interesting questions for researchers. These
systems in general display inherent quantum
interference phenomena which modify the quantum dynamics and produces a
contrast in their evolution.
Quantum interference phenomena, constructive or destructive in nature,
contribute to the quantum recurrences in dynamical systems. Dynamical 
recurrences in such systems originate from the simultaneous excitation of discrete 
quasi-energy states ~\cite{SaifPR}. These recurrences are suitable to use
as a probe to study
quantum chaos \cite{ReinhardtBlumel2001,SaifPRE,Saif2005-a}.
In this contribution we extend the discussion on the quantum recurrence
phenomena and treat
them in higher coupling/strong modulation regimes and study the quantum
dynamics of the system for the nonlinear resonances. 
Our theoretical work is based on Floquet theory \cite{SIChuPR2004}, which presents an elegant formalism
for the analysis of periodically driven and coupled higher dimensional
systems \cite{RobertPRA2002}. Floquet state formalism has been applied to
a number of time-dependent problems: from coherent states of
driven Rydberg atom \cite{VelaPRA2005}, chaotic quantum ratchets
\cite{HurPRA2005}, electron transmission in semiconductor hetero-structures
\cite{ZhangPRB2006}, selectively suppressing of tunneling in quantum-dot array
\cite{VillasPRB2004} to frequency-comb laser fields
\cite{SonPRA2008}. The Floquet states have also been used in Bose-Einstein
condensate systems and applied to probe superfluid-insulator transition
\cite{Eckardt}, towards coherent control \cite{HolthausPRA2001} and dynamical tunneling
\cite{HensingerPRA2004}.

In this paper: (i) We extend Floquet analysis for nonlinear resonances and
explain the existence of robust dynamical recurrences in dynamical system, in
contrast to
delicate dynamical recurrences already reported
\cite{Saif2005-a,SaifPR,Saif2002,ShahidPLA}; (ii) Super revival times in
dynamical systems are discussed first time for both
delicate dynamical recurrences and robust dynamical recurrences; (iii) We show
that,
the non-linearity of the uncoupled systems, and the initial conditions on the
excitation contribute to the classical period, quantum revival time and super
revival time occurring in the coupled higher-dimensional systems or periodically
driven systems;
% We study the phenomena of dynamical recurrences for general systems and
% provide mathematical calculations for recurrence time in the two cases.
(iv) We apply these results to the dynamics of
matter waves in driven optical lattices, a topic of current
research \cite{Applications,KurnNature2008,PottingPRA2001}. Classical dynamics of the
system display an intricate dominant regular and dominant stochastic
dynamics, one
after the other, as a function of increasing modulation amplitude
\cite{Bardroff1995,Raizen1999}.

The paper is organized as follows: The effective Hamiltonians for nonlinear
resonances in higher dimensional systems (Sec. \ref{sec:HDS}) and periodically
driven time dependent systems (\ref{sec:tds}) are derived. In
Sec.~\ref{sec:qrec}, we explain the time scales and their dependence on system
parameters. We apply obtained results to the matter wave dynamics in driven
optical lattices and explain the time scales as a
function of system parameters. We compare analytical and numerical results in
Sec. \ref{sec:NR}.

\section{Floquet Theory of Nonlinear Resonances \label{sec:EHFNR}}

Floquet theory is an elegant formalism to quantize coupled higher dimensional
systems (HDS) and periodically driven system (PDS) \cite{RobertPRA2002}.
In HDS, with two degrees of freedoms coupled, such as Billiards
\cite{Lenz2010} and in
PDS, we observe various dynamical modes. In these Hamiltonian dynamical systems
the stable nonlinear resonances are immersed in stochastic sea and the system
may display global stochasticity beyond a critical value of coupling or
modulation
strength \cite{Lichtenberg,LindaReichl,E.Ott}. Here, we apply
Floquet analysis to find quasi energy eigen states and
quasi energy eigen values in HDS and PDS and discuss their parametric
dependencies on the uncoupled system and coupling
constant or modulation strength.

\subsection{Higher Dimensional Systems \label{sec:HDS}}

We write general Hamiltonian for a system, with its degrees of freedom
coupled, that is,
\begin{equation}
H=H_{0}(I)+\lambda H_{c}(I,\theta),
\end{equation}
where, $H_{0}$ is the Hamiltonian of the system in the absence of coupling,
expressed in the action coordinates $I=(I_{1},I_{2})$. Moreover, $H_{c}$ is
the coupling Hamiltonian which describes coupling and is periodic in angle,
$\theta=(\theta_{1},\theta_{2})$ and the parameter $\lambda$ is coupling
strength. The Hamiltonian may express Billiard systems \cite{Lenz2010} in
general systems and same time is a good candidate to describe multi-atomic
molecule \cite{GoulvenPRA2011} and Lorentz gas \cite{Oliveira} .
We express the
coupling Hamiltonian as,
\begin{equation}
H_{c}=\sum_{n}H_{n}(I)e^{i{n}.\theta}, \label{eq:hsum}%
\end{equation}
where, $n=(n_{1},n_{2})$. Whenever, the frequencies $\Omega=\partial{H_{0}%
}/\partial I$ obey the relation, $n.\Omega=n_{1}\Omega_{1}+n_{2}\Omega_{2}=0$,
resonances occur in the system.

We consider that the coupling exists between actions $I_{1}$ and $I_{2}$.
Within the region of resonance, we find slow variations in action, hence,
following the method of secular perturbation theory, we average over faster
frequency, and get the averaged Hamiltonian for the $N^{th}$ resonance, as
${\bar H}={\bar H}_{0}(I)+\lambda V\cos(N\varphi_{-})$. Here, $\varphi
_{-}=\theta_{1}-(M/N)\theta_{2}$, $I=I_{1}$ is the action corresponding to the
angle $\theta_{1}$, $V$ is the Fourier amplitude, and $M$ and $N$ are
relatively prime integers~\cite{Lichtenberg}. Averaging over rapidly changing
$\varphi_{+}$ makes the corresponding action $I_{2}$ as the constant of
motion. Moreover, ${\bar H}_{0}(I)$ expresses uncoupled averaged Hamiltonian.

The energy of initially narrowly peaked excitation changes slowly when we
produce it in the vicinity of $N^{th}$ non-linear resonance. Therefore, we
expand the unperturbed energy, ${\bar{H}}_{0}(I)$, by means of Taylor
expansion around mean action $I=I_{0}$ and keep only the terms up to second
order~\cite{Flatte}. 

We~\cite{Berry,Almeida} use operator definition of action, that is,
$I=\frac{k^{\hspace{-2.1mm}-}}{i}\frac{\partial}{\partial\theta}$ to get the
near resonant energy spectrum of the system by mapping the corresponding
Schr\"{o}dinger equation on the Mathieu equation \cite{SaifEPJD}, where, the
new variable $\theta=N\varphi/2$, hence, we get
\begin{equation}
\lbrack\frac{\partial^{2}}{\partial{\theta}^{2}}+a-2q\cos2\theta]\phi
(\theta)=0, \label{eq:math}%
\end{equation}
where, the Mathieu characteristic parameter~\cite{AbramowitzStegun,McLachlan},
is
\begin{equation}
a=\frac{8}{N^{2}\zeta{k^{\hspace{-2.1mm}-}}^{2}}[{\mathcal{E}-\bar{H}}%
_{0}+\frac{\omega^{2}}{2\zeta}]. \label{anh}%
\end{equation}
and
\begin{equation}
q=\frac{4\lambda V}{N^{2}\zeta{k^{\hspace{-2.1mm}-}}^{2}}. \label{QH}%
\end{equation}
The parameter $\zeta$ defines
the nonlinear dependence of the energy of the system on the quantum number and
$k^{\hspace{-2.1mm}-}$ is re-scaled Plank's constant.

The $\pi$-periodic solutions of equation~(\ref{eq:math}) correspond to even
functions of the Mathieu equation whose corresponding eigenvalues are
real~\cite{AbramowitzStegun}. These solutions are defined by Floquet states,
\textit{i.e.} $\phi(\theta)=e^{i\mu\theta}P_{\mu}(\theta)$, where, $P_{\mu
}(\theta)=P_{\mu}(\theta+ \pi)$, and $\mu$ is the characteristic exponent. In
order to write the $\pi$-periodic solutions in $\varphi$-coordinate, we
require $\mu$ to be defined as $\mu=\mu(j)=2j/N$, where, $j=0,1,2,....,N-1$.

The allowed values of $\mu(j)$ can exist as a characteristic exponent of
solution to the Mathieu equation for discrete $\nu$ (which takes integer
values) only for certain value $a_{\nu}(\mu(j),q)$, when $q$ is fixed.
Hence, with the help of Eq.~(\ref{anh}), we obtain the values of unknown
$\mathcal{E}$. Therefore, we may express the eigen energy of the system as
\cite{Holthaus1991}
\begin{equation}
\mathcal{E}_{\mu,\nu}=\left[  \frac{N^{2} {k^{\hspace{-2.1mm}-}}^{2}\zeta}%
{8}a_{\nu}(\mu(j),q) +k^{\hspace{-2.1mm}-} \alpha{j}+\bar{H}_{0}%
(I_{0})\right]  \, \mathrm{mod} \,\, k^{\hspace{-2.1mm}-}\omega. \label{ener}%
\end{equation}
Here, $\alpha=M/N$ defines the winding number.

In order to check this result we study the case of zero coupling strength,
that is $\lambda=0$. In this case, the value for Mathieu characteristic
parameter becomes $a_{\nu}(q=0)={\nu}^{2}$. This reduces the quasi-energy
$\mathcal{E}_{\mu,\nu}$, in the absence of coupling
term, that is, $\lambda=0$, provided $\nu$ as $2(I-I_{0})/Nk^{\hspace
{-2.1mm}-}$.

\subsection{Time Dependent Systems}

\label{sec:tds}

In the periodically driven potentials energy is no more a constant of motion.
For the reason we solve the time dependent Schr\"{o}dinger equation by using
secular perturbation approximation as suggested by Max Born~\cite{Born}.
Therefore, the solution is obtained by averaging over rapidly changing
variables. This leads us to find out a partial solution of the periodically
driven systems. As a result, we find quasi energy eigen-states and the quasi
eigen energies of the dynamical system for non-linear resonances.

In order to study the quantum nonlinear resonance of the TDS, we consider the
Hamiltonian,
\begin{equation}
H=H_{0}+\lambda V(x)\sin(t). \label{tdse}%
\end{equation}
The solution of the Schr\"{o}dinger equation corresponding to the
Hamiltonian~(\ref{tdse}) can be written in the form
~\cite{Saif2002,Flatte,GPBerman1977},
\begin{equation}
|\psi(t)\rangle=\sum_{n}C_{n}(t)|n\rangle\exp\{-i[E_{\bar{n}}+(n-\bar{n}%
)\frac{k^{\hspace{-2.1mm}-}}{N}]\frac{t}{k^{\hspace{-2.1mm}-}}\}. \label{psit}%
\end{equation}
Here, $E_{\bar{n}}$ is the mean energy, $C_{n}(t)$ is time dependent
probability amplitude, $\bar{n}$ is mean quantum number and $|n\rangle$ are
eigen states of undriven system.  
On substituting equation~(\ref{psit}) in the time dependent Schr\"{o}dinger
equation, we find that the probability amplitude $C_{n}(t),$ changes with time
following the equation, $ik^{\hspace{-2.1mm}-}{\dot{C}}_{n}(t)=[E_{n}%
-E_{\bar{n}}-(n-\bar{n})\frac{k^{\hspace{-2.1mm}-}}{N}]C_{n}(t)+\frac{\lambda
V}{2i}(C_{n+N}-C_{n-N}).$ Where, $V=V_{n-N}\approx V_{n+N}$ are off-diagonal
matrix elements and are approximately constant near the potential minima for
tight-binding approximations.

We take the initial excitation, narrowly peaked around the mean value,
$\bar{n}$. For the reason, we take slow variations in the energy, $E_{n}$,
around the $\bar{n}$ in a nonlinear resonance, and expand it up to second
order in Taylor expansion. Thus, the Schr\"{o}dinger equation for the
probability amplitudes, $C_{n}(t)$, is
\begin{align}
ik^{\hspace{-2.1mm}-}{{\dot{C}}_{n}}  &  =k^{\hspace{-2.1mm}-}(n-\bar
{n})(\omega-\frac{1}{N})C_{n}(t)+\frac{1}{2}k^{\hspace{-2.1mm}-}(n-\bar
{n})^{2}\zeta C_{n}(t)\nonumber\\
&  +\frac{\lambda V}{2i}(C_{n+N}-C_{n-N}). \label{prob}%
\end{align}
In order to obtain equation~(\ref{prob}), we average over the fast oscillating
terms and keep only the resonant ones. Here, the frequency, $\omega=\frac{\partial E_{n}%
}{k^{\hspace{-2.1mm}-}\partial n}|_{n=\bar{n}},$ and the non-linearity, in the
time independent system is given as $\zeta=\frac{\partial^{2}E_{n}}%
{k^{\hspace{-2.1mm}-2}\partial n^{2}}|_{n=\bar{n}}.$

We introduce the Fourier representation for $C_{n}(t)$ as,
\begin{equation}
C_{n}(t)=\frac{1}{2N\pi}\int_{0}^{2N\pi}g(\theta,t)e^{-i(n-\bar{n})\theta
/N}d\theta, \label{four}%
\end{equation}
which helps us to express equation~(\ref{prob}) as the Schr\"{o}dinger
equation for $g(\theta,t)$, such that $ik^{\hspace{-2.1mm}-}{\dot{g}}%
(\theta,t)=H(\theta)g(\theta,t)$. Here, the Hamiltonian $H(\theta)$ is given
as, $H(\theta)=-\frac{N^{2}k^{\hspace{-2.1mm}-2}\zeta}{2}\frac{\partial^{2}%
}{\partial\theta^{2}}-iNk^{\hspace{-2.1mm}-}(\omega-\frac{1}{N})\frac
{\partial}{\partial\theta}-\lambda V\sin\theta$. In order to obtain this
equation, we consider the function $g(\theta,t)$ as $2N\pi$ periodic, in
$\theta$ coordinate.

Due to the time-independent behavior of the Hamiltonian, we write the time
evolution of $g(\theta,t)$, as $g(\theta,t)=\tilde{g}(\theta)e^{\frac
{-i\varepsilon t}{k^{\hspace{-2.1mm}-}}}$. Therefore, Schr\"{o}dinger equation
for $g(\theta,t)$ reduces to the standard Mathieu equation (\ref{eq:math}).
Here, we express $\tilde{g}=\chi(z)\exp\left(  -i2(N\omega-1)z/N^{2}\zeta
k^{\hspace{-2.1mm}-}\right)  $, where, $\theta=2z+\pi/2$. The same as in
equation~(\ref{eq:math}), the Mathieu characteristic parameter, $a$, is
defined as
\begin{equation}
a=\frac{8}{N^{2}k^{\hspace{-2.1mm}-2}\zeta}[\frac{(N\omega-1)^{2}}{2N\zeta
}+\mathcal{E_{\mu,\nu}}], \label{anew}%
\end{equation}%
\begin{equation}
\text{and \ \ \ \ \ \ \ }q=\frac{4\lambda V}{N^{2}k^{\hspace{-2.1mm}-2}\zeta}.
\label{qq}%
\end{equation}
Hence, the quasi eigen energy of the system, obtained from
equation~(\ref{anew}), is
\begin{equation}
\mathcal{E}_{\mu,\nu}=\left[ \frac{N^{2}k^{\hspace{-2.1mm}-2}\zeta}{8}a_{\nu
}(\mu(j),q)+k^{\hspace{-2.1mm}-}\alpha{j} \right] \,\, \mathrm{mod}
\,\,k^{\hspace{-2.1mm}-}\omega, \label{ener1}%
\end{equation}
where, the index $\nu$ takes the definition, $\nu=\frac{2(n-\bar{n})}{N}$,
where, $n$ is the quantum number for time independent system.

\section{Quantum Revival Times Based on Quasi Energy Spectrum \label{sec:qrec}}

In this section we discuss two cases correspond to nonlinear
resonances. We consider weakly coupled $q<1$ or strongly coupled  
$q\gg1$ potentials. In $q<1$ situation, for large $\nu$ \cite{Saif2002}
and in $q\gg1$ situation for small $\nu$, near the center of resonance matrix
elements are constant.
In the following discussion, we analyze the wave packet dynamics in these
regimes and show their parametric dependencies.

The time scales, $T^{(j)}$ at which recurrences of an initially well-localized
wave packet occur depend on the quasi-energy of the respective system, The
recurrence times are obtained as, $T^{(j)}=2\pi/\Omega^{(j)}$, hence, the
values of $j$ as $j=1,2,3...$, correspond to, respectively, classical,
quantum, super, and higher order revival times. With the help of
Eqs.~(\ref{ener}) and (\ref{ener1}), we obtain the frequencies $\Omega^{(j)}$
as
\begin{align}
\Omega^{(1)}  & =\frac{1}{k^{\hspace{-2.1mm}-}}\left\{  \frac{\partial
\mathcal{E}_{\mu,\nu}}{\partial\mu}+\frac{\partial\mathcal{E}_{\mu,\nu}%
}{\partial\nu}\right\}  \nonumber\\
\Omega^{(2)}  & =\frac{1}{2!k^{\hspace{-2.1mm}-2}}\left\{  \frac{\partial
^{2}\mathcal{E}_{\mu,\nu}}{\partial\mu^{2}}+\frac{2\partial^{2}\mathcal{E}%
_{\mu,\nu}}{\partial\mu\partial\nu}+\frac{\partial^{2}\mathcal{E}_{\mu,\nu}%
}{\partial\nu^{2}}\right\}  \label{freq}\\
\Omega^{(3)}  & =\frac{1}{3!k^{\hspace{-2.1mm}-3}}\left\{  \frac{\partial
^{3}\mathcal{E}_{\mu,\nu}}{\partial\mu^{3}}+3\frac{\partial^{3}\mathcal{E}%
_{\mu,\nu}}{\partial\mu^{2}\partial\nu}+3\frac{\partial^{3}\mathcal{E}%
_{\mu,\nu}}{\partial\mu\partial\nu^{2}}+\frac{\partial^{3}\mathcal{E}_{\mu
,\nu}}{\partial\nu^{3}}\right\}  .\nonumber
\end{align}

After a few mathematical steps Eqs.~(\ref{freq}) reduce to
\begin{align}
\Omega^{(1)}  & =\frac{1}{k^{\hspace{-2.1mm}-}}\left\{  \frac{\partial
\mathcal{E}_{\mu,\nu}}{\partial\nu}+\alpha k^{\hspace{-2.1mm}-}\right\}
\nonumber\\
\Omega^{(2)}  & =\frac{1}{2!k^{\hspace{-2.1mm}-2}}\left\{  \frac{\partial
^{2}\mathcal{E}_{\mu,\nu}}{\partial\nu^{2}}+\alpha k^{\hspace{-2.1mm}-}%
\frac{2\partial\mathcal{E}_{\mu,\nu}}{\partial\nu}\right\} \nonumber \\
\Omega^{(3)}  & =\frac{1}{3!k^{\hspace{-2.1mm}-3}}\left\{  \frac{\partial
^{3}\mathcal{E}_{\mu,\nu}}{\partial\nu^{3}}+3\alpha k^{\hspace{-2.1mm}-}%
\frac{\partial^{2}\mathcal{E}_{\mu,\nu}}{\partial\nu^{2}}\right\}\label{freq1}
\end{align}
which lead to
%where, $\omega^{(d)}=(j!k^{\hspace
%{-2.1mm}-})^{-1}\partial^{d}\varepsilon_{\nu}/\partial{\nu}^{j}$
%\cite{Saif2005-a,SaifEPJD}.
classical period, $T^{(1)}=T_{\lambda}^{(cl)},$ quantum revival time,
$T^{(2)}=T_{\lambda}^{(rev)},$ and super revival time, $T^{(3)}=T_{\lambda
}^{(spr)}$, when calculated at the mean values.

\subsubsection{Delicate Dynamical Recurrences}

The condition, $q<1$, may be satisfied in the presence of weak
perturbation due to external periodic force (and/or), for large nonlinearity
and/or
for large effective Plank's constant
\cite{Saif2002,Saif2005-a,SaifPR,ShahidPLA}. The Mathieu characteristic
parameters, $a_{\nu}$ and $b_{\nu}$\ are given \cite{AbramowitzStegun}, as%

\begin{align}
a_{\nu}  &  \simeq b_{\nu}=\nu^{2}+\frac{q^{2}}{2(\nu^{2}-1)}+...\text{. for
}\nu\geqslant5 \label{shallowA}%
\end{align}

The above expressions are not limited to integral value of $\nu$ and are very
good approximations when $\nu$ is of the form, $m+\frac{1}{2}$. In case of
integral value of $\nu$, i.e. $\nu=m$, the series holds only up to the terms
not involving $\nu^{2}-m^{2}$\ in the denominator.

The energy spectrum for weakly modulated periodic potentials \cite{SaifEPJD}
can be defined using equations (\ref{ener1}) and (\ref{shallowA}). The
relations obtained for classical period, quantum revival time and super
revival time, in the presence of small perturbation for primary resonance $N=1$,
index $j$ takes the value, $j=0$, and time scale are simplified as 
\begin{equation}
T_{\lambda}^{(cl)}=(1-M^{(cl)})T_{0}^{(cl)}\Delta,\label{TShallowClGen}%
\end{equation}%
\begin{equation}
T_{\lambda}^{(rev)}=(1-M^{(rev)})T_{0}^{(rev)},\label{TShallowRevGen}%
\end{equation}%
\begin{equation}
T_{\lambda}^{(spr)}=\frac{\pi\omega^{2}}{2\lambda V\zeta\Delta^{2}}%
\frac{(1-\mu_1^{2})^{4}}{\mu_1},\label{TshallowSupGen}%
\end{equation}
here, $T_{0}^{(cl)}=\frac{2\pi}{\omega}$ is classical period and
$T_{0}^{(rev)}=2\pi/(\frac{1}{2!}k^{\hspace{-2.1mm}-}\zeta)$ is quantum
revival time in the absence of external modulation. Furthermore,
$\Delta=(1-\frac{\omega_{N}}{\omega})$. The modification factors are given as
\[
M^{(cl)}=-\frac{1}{2}(\frac{\lambda V\zeta\Delta^{2}}{\omega^{2}}%
)^{2}\frac{1}{(1-\mu_1^{2})^{2}}%
\]
and%
\[
M^{(rev)}=\frac{1}{2}(\frac{\lambda V\zeta\Delta^{2}}{\omega^{2}}%
)^{2}\frac{3+\mu_1^{2}}{(1-\mu_1^{2})^{3}},
\]
here, $\mu_1=\frac{k^{\hspace{-2.1mm}-}\zeta\Delta}{2\omega}$ is re-scaled
non-linearity.

\subsubsection{Robust Dynamical Recurrences}

The condition, $q\gg1$, may be satisfied in the presence of large amplitude,
$\lambda$, of external modulation by periodic force in a dynamical system. In
addition, we may get the regime by considering a system with very small
linearity i.e. $\zeta \approx 0$ and/or by taking effective Plank's constant
$k^{\hspace{-2.1mm}-}$ be very small. Quasi energies for nonlinear
resonance are defined in terms of Mathieu characteristic parameters, which are%
\begin{equation}
a_{\nu}\approx b_{\nu+1}\approx-2q+2s\sqrt{q}-\frac{s^{2}+1}{2^{3}}%
-\frac{s^{3}+3s}{2^{7}\sqrt{q}}-..., \label{Deep A}%
\end{equation}
where, $s=2\nu+1.$

The energy spectrum of nonlinear resonances for strongly modulated periodic
potentials can be defined using equations (\ref{ener1}) and (\ref{Deep A}).
Here, in the deep potential limit, the band width is,
\begin{equation}
b_{\nu+1}-a_{\nu}\simeq\frac{2^{4\nu+5}\sqrt{\frac{2}{\pi}}q^{\frac{\nu}%
{2}+\frac{3}{4}}\exp(-4\sqrt{q})}{\nu!}.
\end{equation}

Keeping lower order terms in $s$, in equation (\ref{Deep A}), we get quasi
energy spectrum for nonlinear resonances as $(\nu+\frac{1}{2})${$k^{\hspace
{-2.1mm}-}$}$\omega_{h}$, which resembles to the harmonic oscillator energy
spectrum for $\omega_{h}=2\sqrt{V_{0}}$.

Similarly the relations for classical period, quantum revival time and super
revival time for strongly driven case are obtained as,

\begin{equation}
T_{\lambda }^{(cl)}=\frac{4\pi }{N^{2}k^{\hspace{-2.1mm}-}\zeta \sqrt{q}%
\left\{ 1-\frac{2v+1}{8\sqrt{q}}-\frac{3(2v+1)^{2}+3}{2^{8}q}\right\}
+2\alpha }
\end{equation}%
\begin{equation}
T_{\lambda }^{(rev)}=\frac{32\pi }{N^{2}\zeta \left\{ 1+\frac{3(2v+1)}{16%
\sqrt{q}}+8\alpha k^{\hspace{-2.1mm}-}\sqrt{q}-(2v+1)\ \alpha k^{\hspace{%
-2.1mm}-}\right\} }
\end{equation}%
\begin{equation}
T_{\lambda }^{(spr)}=\frac{32\pi }{N^{2}\zeta \alpha }[1-\frac{1}{8\alpha k^{%
\hspace{-2.1mm}-}\sqrt{q}}\{1+\ \frac{3\alpha k^{\hspace{-2.1mm}-}(2v+1)}{2}%
\}].
\end{equation}

For primary resonance $N=1$, index $j$ takes the value, $j=0$, and time scale
are simplified as

\begin{equation}
T_{\lambda}^{(cl)}=T_{0}^{(cl)}\frac{\Delta}{8\mu_1}\left[ 1-\frac{(4+\mu_1
)\sqrt{\zeta}}{8\mu_1\sqrt{q}}-\frac{(4+\mu_1)^{2}\zeta}{(8\mu_1)^{2} q}\right],
\label{TdeepClGen}%
\end{equation}%
\begin{equation}
T_{\lambda}^{(rev)}=T_{0}^{(rev)}2 \left[ 1-\frac
{3(\mu_1+4)}{16\mu_1\sqrt{q}}+\frac{9(4+\mu_1)^{2}}{(16\mu_1)^{2} q}\right] ,
\label{TdeepRevGen}%
\end{equation}
and%
\begin{equation}
T_{\lambda}^{(spr)}=\frac{2^{5}\pi\sqrt{q}}{k^{\hspace{-2.1mm}-}\zeta}.
\label{TdeepSupGen}%
\end{equation}
{\it Atoms in modulated optical lattice:}
In order to analyze the time scales in equations
(\ref{TShallowClGen})-(\ref{TshallowSupGen}) 
and equations (\ref{TdeepClGen}-\ref{TdeepSupGen}), we consider ultracold
Rubidium atoms, 
$Rb^{87}$, in standing wave field with
phase modulation due to acousto-optic modulator. The system is of immense
theoretical and experimental interest since last two decades in atom optics
with cold atoms \cite{Raizen1999} and Bose-Einstein condensate
\cite{Yukalov2009}. The atoms
in the presence of phase modulation experience an external force, thus exhibit
dispersion both in classical and quantum domain. In quantum dynamics atoms
experience an additional control due to effective Plank's constant. This as
reported earlier limits the classical evolution observed in classical
prototype and results in dynamical localization \cite{Moore1994}. Here, we
report that in long time evolution, material wave packet display quantum
recurrence phenomena. The dynamics of cold atoms in phase modulated optical
lattice is governed by the Hamiltonian, which is%
\[
H=\frac{p^{2}}{2M}+\frac{V_{0}}{2}\cos[2k_{L}\{x-\Delta L\sin(\omega
_{m}t)\}],
\]
where, $k_{L}$ is wave number and $V_{0}$ define the potential depth of an
optical lattice. Furthermore, $\Delta L$ and $\omega_{m}$ are amplitude and
frequency, whereas, $M$ is the mass of an atom moving in the optical lattice.

The unitary transformations\footnote{The unitary, time
periodic transformations preserves the quasi-energy spectrum.}\\
$\psi(x,t)=\tilde{\psi
}(x,t)\exp[\frac{i}{\hbar}\{\frac{\omega_{m}M\Delta L\cos(\omega_{m}t)}%
{2}x+\beta(t)\}],$ where, $\beta(t)=\frac{\omega_{m}^{2}\Delta L^{2}M}%
{4}[\frac{\sin(2\omega_{m}t)}{2\omega_{m}}+t],$ to a frame co-moving with the
lattice, modified the Hamiltonian as
\begin{equation}
H=\frac{p^{2}}{2M}+\frac{V_{0}}{2}\cos2k_{L}x+Fx\sin\omega_{m}t,
\label{ModHam}%
\end{equation}
where, $F=M\Delta L\omega_{m}^{2}$ is amplitude of inertial force emerging in
the oscillating frame. In laboratory accessible quantities, scaled optical
lattice depth $\frac{V_{0}}{E_{r}}$ is $5$ to $20$. For example $Rb^{87}$
atoms trapped by optical lattice of wavelength $842nm$ has $E_{r}%
=1.34\mathrm{x}10^{-11}eV$, thus recoil energies are of the order of
$10^{-10}eV$ \cite{Holthaus2010}. To examine the dynamics of cold atoms in
driven optical lattices, Hamiltonian (\ref{ModHam}) is expressed in
dimensionless quantities. Using scaling transformations $z=k_{L}%
x,\ \tau=\omega_{m}t$, $\Psi(z,\tau)=\tilde{\psi}(x,t)\ $ and multiplying the
Schr\"{o}dinger wave equation by $\frac{2\omega_{r}}{\hbar\omega_{m}^{2}},$
where, $\omega_{r}=\frac{\hbar k_{L}^{2}}{2M},$ is single photon recoil
frequency, we get dimensionless Hamiltonian,%
\[
\tilde{H}=-\frac{k^{\hspace{-2.1mm}-2}}{2}\frac{\partial^{2}}{\partial z^{2}%
}+\frac{\tilde{V}_{0}}{2}\cos2z+\lambda z\sin\tau,
\]
here, $\lambda=k_{L}\Delta L,$ is modulation amplitude, i.e. the effective
amplitude of inertial force, $\tilde{V}_{0}=\frac{V_{0}k^{\hspace{-2.1mm}-}%
}{\hbar\omega_{m}}$ and $\tau$ is scaled time in the units of modulation
frequency $\omega_{m}$. In this case the rescaled Plank's constant
$k^{\hspace{-2.1mm}-}=\frac{2\omega_{r}}{\omega_{m}}.$

As time scales of driven optical lattice are expressed in terms of undriven
system parameters, for the reason, we investigate, classical period, quantum
revival time and super revival time scales of undriven optical lattice
\cite{DrytingPRA1993,AyubJRLR2009,Holthaus1997}.

In the case of undriven shallow optical potential $T_{0}^{(cl)}=(1+\frac
{q_{0}^{2}}{2(n^{2}-1)^{2}})\frac{\pi}{n},$ $T_{0}^{(rev)}=2\pi(1-\frac
{q_{0}^{2}}{2}\frac{(3n^{2}+1)}{(n^{2}-1)^{3}})$ and $T_{0}^{(spr)}=\frac
{\pi(n^{2}-1)^{4}}{q_{0}^{2}n(n^{2}+1)}$, where, $n$ is band index of undriven
lattice, while, $q_{0}=\frac{V_{0}}{4\omega_{r}}$ is rescaled potential depth
in units of single recoil energy  $E_{r}=\frac{\hbar^2 k_{L}^{2}}{2M}$. 
On the other hand for deep optical lattice, the
classical time period is $T_{0}^{(cl)}=\frac{\pi}{2\sqrt{q_{0}}}(1+\frac
{s}{8\sqrt{q_{0}}}+\frac{3(s^{2}+1)}{2^{8}q_{0}}),$ where, $s=2n+1$. The
quantum revival time $T_{0}^{(rev)}=4\pi(1-\frac{3s}{16q_{0}})$ and super
revival time is $T_{0}^{(spr)}=32\pi\sqrt{q_{0}}$ \cite{AyubJRLR2009}.

In shallow lattice potential limit, i.e. $q_{0}<1$, neglecting the
higher order terms in $q_{0}$ the classical frequency, $\omega=2n(1-\frac
{q_{0}^{2}}{2(n^{2}-1)^{2}})$ and non-linearity, $\zeta=2+\frac{q_{0}^{2}}%
{2}\frac{3n^{2}+1}{(n^{2}-1)^{3}}$. In deep optical potential limiting case,
i.e. $q_{0}\gg1$, the classical frequency is $\omega=4(\sqrt{q_{0}}-\frac
{2n+1}{8})$ and non-linearity is $\zeta=|-1-\frac{3(2n+1)}{2^{4}\sqrt{q_{0}}%
}|$.

For driven optical potential, keeping in view the periodicity of Floquet
solutions \cite{Saif2002,Ayub2010}, we take only even values of the index
\cite{AbramowitzStegun}. Hence, re-scaled Mathieu characteristic exponent,
$\nu=2(l+\beta), \text{ where, } \beta=\frac{N\omega-1}{N^{2}\zeta
k^{\hspace{-2.1mm}-}}$.

Comparing the coefficients of eigen energy of the undriven system and equation
of motion of probability amplitude in the absence of modulation, we get
$l=\frac{n-n_{0}}{N},$ which is new band index for nonlinear resonance and at
the center of resonance $l=0$. 

For delicate dynamical recurrences, time scales for primary resonance
$N=1$, are
\begin{equation}
T_{\lambda}^{(cl)}=T_{0}^{(cl)}\left[ 1+\frac{q^{2}}{2}\frac{1}{\{4(l+\beta
)^{2}-1\}^{2}}\right] \Delta, \label{TshallowCl}%
\end{equation}%
\begin{equation}
T_{\lambda}^{(rev)}=T_{0}^{(rev)}\left[ 1-\frac{q^{2}}{2}\frac{12(l+\beta
)^{2}+1}{\{4(l+\beta)^{2}-1\}^{3}}\right] , \label{TshallowRev}%
\end{equation}%
\begin{equation}
\text{and \ \ \ }T_{\lambda}^{(spr)}=\frac{\pi\{4(l+\beta)^{2}-1\}^{4}}%
{2N^{2}\zeta k^{\hspace{-2.1mm}-}q^{2}(l+\beta)\{4(l+\beta)^{2}+1\}}.
\label{TshallowSup}%
\end{equation}
Where, $T_{0}^{(cl)}=\frac{2\pi}{\omega(l+\beta)}$ is the classical time
period and $T_{0}^{(rev)}=$ $\frac{4\pi}{k^{\hspace{-2.1mm}-}\zeta}$ is
quantum revival time for unmodulated system.

On the other hand, in deep optical lattice limiting case, time scales for the
atomic wave packet for primary resonance with $N=1$, are given as%
\begin{equation}
T_{\lambda}^{(cl)}=\frac{2\pi}{N^{2}k^{\hspace{-2.1mm}-}\zeta\{\sqrt{q}%
-\dfrac{4(l+\beta)+1}{8\ }\}}, \label{TdeepCl}%
\end{equation}%
\begin{equation}
T_{\lambda}^{(rev)}=\frac{8\pi}{N^{2}k^{\hspace{-2.1mm}-}\zeta}\left[
1-\dfrac{3\{4(l+\beta)+1\}}{16\sqrt{q}}\right]  \label{TdeepRev}%
\end{equation}
and%
\begin{equation}
T_{\lambda}^{(spr)}=\frac{32\pi\ \sqrt{q}}{N^{2}k^{\hspace{-2.1mm}-}\zeta}.
\label{TdeepSup}%
\end{equation}

In case of deep lattice, when external modulation frequency is close to the
harmonic frequency, matrix elements, $V$ can be approximated by those of
harmonic oscillator and Mathieu parameter $q$\ can be approximated as
$q\approx\frac{4\sqrt{n+1}\lambda}{q_{0}^{\frac{1}{4}}k^{\hspace{-2.1mm}%
-2}\zeta}$ \cite{Holthaus1997} . Under this approximation time scales are%
\begin{equation}
T_{\lambda}^{(cl)}=\frac{16\pi q_{0}^{\frac{1}{8}} \text{/} N^{2}\sqrt{\zeta}%
}{16(n+1)^{\frac{1}{4}}\sqrt{\lambda}-\{4(l+\beta)+1\}q_{0}^{\frac{1}{8}%
}k^{\hspace{-2.1mm}-}\sqrt{\zeta}}, \label{TdeepClHar}%
\end{equation}

\[
T_{\lambda}^{(rev)}=\frac{8\pi}{N^{2}k^{\hspace{-2.1mm}-}\zeta}\left[
1-\frac{3\{4(l+\beta)+1\}q_{0}^{\frac{1}{8}}k^{\hspace{-2.1mm}-}\sqrt{\zeta}%
}{32(n+1)^{\frac{1}{4}}\sqrt{\lambda}}\right]
\]

\begin{equation}
\text{and }T_{\lambda}^{(spr)}=\frac{64\pi(n+1)^{\frac{1}{4}}\sqrt{\lambda}%
}{N^{2}k^{\hspace{-2.1mm}-2}\zeta^{\frac{3}{2}}q_{0}^{\frac{1}{8}}}.
\label{TdeepSprHar}%
\end{equation}

Behavior of classical periods, quantum revivals and super revivals of matter
waves in modulated optical crystal in nonlinear resonances versus modulation is
shown in Fig-\ref{fig:recure}. 
In each plot of this figure, left vertical axis shows the
time scales when shallow optical lattice is weakly or strongly modulated and
right
axis shows the time scales when deep optical lattice is weakly or strongly
modulated.
The upper row of Fig-\ref{fig:recure} represents the time scales for small $q$
values i.e. delicate dynamical recurrences as a function of
modulation $\lambda$, while, 
lower row represents the time scales when $q\gg1$, as a function of modulation
$\lambda$, i.e. robust dynamical recurrences. Here, left
column 
shows the results related to the classical periods. Quantum revival times 
are plotted in middle column, while, right column shows super revival times.
    
We note that  when optical
lattice is perturbed by weak periodic force, the classical period increases with
modulation, as given in equation (\ref{TshallowCl}). Classical period for weakly
driven shallow lattice potential changes slowly as compared to weakly driven
deep lattice potential as shown in Fig-\ref{fig:recure}(a).
When lattice is strongly modulated by an external periodic force, the
classical period decreases as modulation increases. Classical period for
strongly driven optical lattice is given by equation (\ref{TdeepCl}). The
behavior of classical period for strongly driven lattice versus modulation is
qualitatively of the same order for both strongly driven shallow lattice and
strongly driven deep lattice as shown in Fig-\ref{fig:recure}(b). The behavior
of classical period in strongly driven lattice case is understandable as
strong modulation influence more energy bands of undriven lattice to
follow the external frequency and near the center of nonlinear resonance the
energy spectrum is almost linear, as can be inferred from equations
(\ref{ener1}) and (\ref{Deep A}) with assumptions $q\gg1$ and $l$ is
small.

Quantum revival time in nonlinear resonances versus modulation is shown in
middle column of Fig-\ref{fig:recure}. For delicate dynamical recurrences
Fig-\ref{fig:recure}(c),  the
quantum revival time decreases as modulation increases. The behavior of
quantum revival time is given by equation
(\ref{TshallowRev}) and equation (\ref{TdeepRev}) respectively for delicate
dynamical recurrence and robust dynamical recurrence. For weakly driven
shallow optical lattice or weakly driven deep lattice, qualitative and
quantitative
behavior of revival time is almost similar. The qualitative behavior of
revival time for strongly driven shallow lattice is different from the
strongly driven deep lattice Fig-\ref{fig:recure}(d), as in the later case,
change in revival time is
almost one order of magnitude larger than the former case, for equal
changes in
modulation. Classical period and quantum revival time for delicate dynamical
recurrences show good numerical and analytical resemblance for the
system with our previous work \cite{Saif2005-a,ShahidPLA,SaifEPJD}. Here, the
difference in the
revival time
behavior for strongly driven shallow lattice as compared to strongly driven
deep lattice is due to the difference in energy spectrum of undriven system.
In deep lattice, due to small non-linearity more energy bands are influenced by
the external drive and resonance spectrum is similar to that of harmonic
oscillator near the
center of nonlinear resonance. As modulation is
increased more and more energy bands are influenced by external drive.
Fig-\ref{fig:Spatiotemp} shows spatio-temporal evolution of an initially well
localized wave
packet in a lattice minima. Upper panel is for the
spatio-temporal dynamics of atomic wave packet
in the absence of periodic modulation, while the lower panel presents the case
when external modulation, $\lambda=3$, $q=85.14$. Time evolution of wave packet
in optical lattice shows that wave packet splits into small
wavelets and spreads over the neighboring lattice sites. Later, these wavelets
constructively interfere and wave packet revival takes place. It is clear that
with the introduction of modulation, revival time changes as shown in
Fig-\ref{fig:recure} which is due to a change in interference pattern and
confirms analytical results as discussed above.

The super revival time behavior versus modulation, shown in right column of 
Fig-\ref{fig:recure}, for delicate dynamical recurrences and robust dynamical
recurrences are given respectively in equations
(\ref{TshallowSup}) and (\ref{TdeepSup}). As lattice is slightly perturbed by
periodic external force, the super revival time, $T_{\lambda}^{(spr)}$ decreases
with modulation for delicate dynamical recurrences as shown in
Fig-\ref{fig:recure}(e). On the other hand, the super revival for
robust dynamical recurrences increases with modulation as shown in the
Fig-\ref{fig:recure}(f). Here, qualitative behavior of super revival time is
same but quantitatively super revival time increases almost two times faster.

\section{Discussion}\label{sec:NR}
We presented general discussion on the occurrence of robust
dynamical recurrences  in higher dimensional and time periodic systems in
the presence of strong coupling/modulation.
The dependence on energy spectrum of time scales is
explained. We applied our results to the atomic dynamics in
optical lattice driven by the periodic forcing. 

We have two cases for each condition i.e. for $q<1$
condition which is satisfied either shallow or deep potential is slightly
perturbed by small external periodic force. Analytical expressions
(\ref{TShallowClGen}-\ref{TshallowSupGen}) for wave packet time scales are
valid. Parameters, $V$, $\omega$ and $\zeta$
are scaled matrix elements,
frequency and non-linearity of respective undriven shallow or deep potential.
Similarly, condition $q\gg1$ is satisfied when shallow or deep potential is
strongly modulated. Expressions (\ref{TdeepClGen}-\ref{TdeepSupGen}) represent
time scales of the driven system in this case.
Moreover, for deep lattice case (undriven or driven), expansion
(\ref{Deep A}) is quite good for Mathieu characteristic exponent $\nu$
satisfying the condition $\nu\ll\sqrt{q}$ and for $\nu\gg\sqrt{q}$, equation
(\ref{shallowA}), satisfy the numerically obtained energy spectrum. The
intermediate range ($\nu\sim\sqrt{q}$), where, energy spectrum changes its
character from lower to high value is estimated as $\nu_{c}\approx
2\parallel\sqrt{\frac{q}{2}}\parallel$ \cite{ReyPRA2005}, here,
$\shortparallel y\shortparallel$ denotes the closest integer to $y$.

We numerically explore the dynamics of quantum particle in a modulated
optical lattice for a nonlinear resonance. Numerical results are obtained
by placing wave packet
 in a primary resonance with $N=1$. This is realized in experimental setup of
Mark Raizen at Austin, Texas. The authors
\cite{Moore1994} worked with \ sodium atoms to observe the quantum mechanical
suppression of classical diffusive motion and employed the $(3S_{\frac{1}{2}%
},F=2)\rightarrow(3P_{\frac{3}{2}},F=3)$ transition at $589nm$ , with
$\omega_{0}/2\pi=5.09\times10^{14}Hz.$ The detuning was $\delta_{L}%
/2\pi=5.4\times10^{9}Hz.$ The recoil frequency of sodium atoms was $\omega
_{r}/2\pi=25kHz$ for selected laser frequency and modulation frequency was
chosen $\omega_{m}/2\pi=1.3MHz$, whereas, the other parameters were
$k^{\hspace{-2.1mm}-}=0.038$ and $q=55$ (or $\tilde{V}_{0}=0.16$). In the
other experiment \cite{Ben1996}, Bloch oscillations of ultracold atoms were
observed with the $6S_{\frac{1}{2}}\rightarrow6P_{\frac{3}{2}}$ transition in
cesium atoms setting $\lambda_{L}=852nm$ and $\omega_{0}/2\pi=3.52\times
10^{14}Hz$. The detuning was $\delta_{L}/2\pi=3\times10^{10}Hz,$ with $q$ up
to $1.5$. The recoil frequency was $\omega_{r}/2\pi=2.07kHz$, so that a
driving frequency $\omega_{m}/2\pi=10^{3}Hz$, three orders of magnitude lower
than in the former experiment, gives $k^{\hspace{-2.1mm}-}\approx4$, taking
the dynamics to the deep quantum regime. We numerically investigate the
validity of our results with known theoretical
\cite{Bardroff1995} and experimental \cite{Moore1994}
results with dimensionless rescaled Plank's constant $k^{\hspace{-2.1mm}%
-}=0.16$, $\tilde{V}_{0}=0.36,$ which is a case of strong modulation to a deep
optical
lattice of potential depth $V_{0}=28.13 E_{r}$. The wave packet is initially
well localized in such a way that localization length is less than or order of
lattice spacing.

To have an idea about the classical dynamics of the system, we plot the
Poincar\'e surface of section for modulation strengths $\ \lambda=0, 1.5$ and
$3.14$ as shown in Fig-\ref{fig:FigPhasespace}. From phase space plot, we see
the appearance of 1:1 resonance for $\lambda>0$. This resonance emerges when
the time period of external force matches with period of unperturbed system. One
effect
of an external modulation is the development of stochastic region near the
separatrix. As modulation is increased, while the frequency is fixed, the size
of stochastic region increases at the cost of regular region.

In order to observe the dynamics of a quantum particle inside a resonance, we
evolve a well localized Gaussian wave packet in the driven optical
lattice. Square of auto-correlation function (${\mid C\mid}^{2}$) for the
minimum uncertainty wave packet is plotted as a function of time in
Fig-\ref{fig:Auto15}, for $\lambda=1.5, q=27.78$, whereas, $\Delta
p=\Delta z=0.5$, $k^{\hspace{-2.1mm}-}=0.5$ and $V_{0}=16 E_r$. In
Fig-\ref{fig:Auto15} upper panel is an enlarged view which displays classical
periods, while middle panel of the
figure shows quantum revivals. Lower panel shows the existence of super
revivals. The classical period, quantum
revival time and super revival time are indicated by arrows and showing there
characteristics. Numerical results are in good agreement with analytical expressions.

\section{Acknowledgment}

M. A. Thanks HEC Pakistan for financial support through grant
no. 17-1-1(Q.A.U)HEC/Sch/2004/5681. F. S. thanks HEC Pakistan for partial
financial support under NRP-20-1374 and Pr\'o-Reitoria de Pesquisa-UNESP,
Brazil. F. S. thanks E D Leonel for his hospitality at DEMAC, UNESP where a part
of this work is completed.
% For one-column wide figures use

\begin{onecolumn}

\begin{figure}[p]
\includegraphics{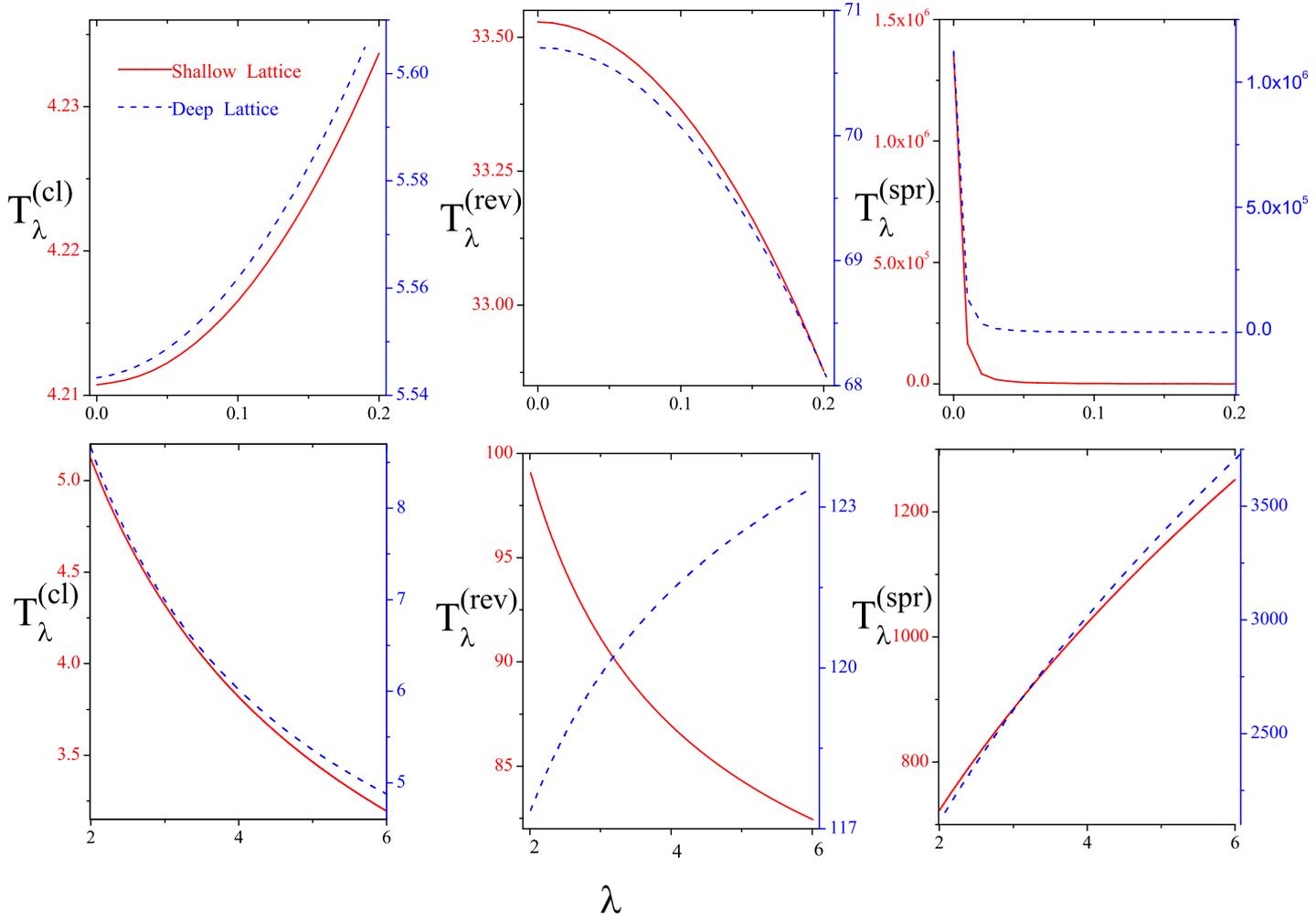}

\caption{Left panel: Classical time period versus $\lambda$ for weak external
modulation (a) and for strong external modulation (b). In this figure, for deep
lattice $V_{0}=16E_r$,  for shallow lattice $V_{0}=2E_r$ and
$k^{\hspace{-2.1mm}-}=0.5$. Middle panel: Quantum revival time versus $\lambda$
for weak external modulation (c) and for strong external modulation (d). Right
panel: Super revival time versus $\lambda$ when external modulation is weak
(e) and for strong external modulation (f).}%
\label{fig:recure}%
\end{figure}
\begin{figure}[p]
\includegraphics{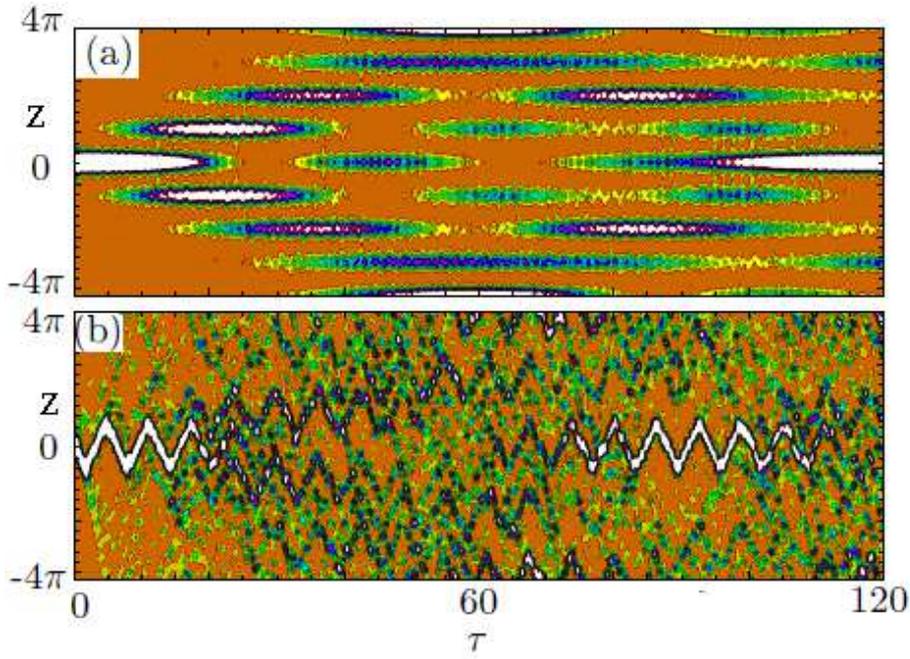}
%{./graphics/new3__7.pdf}
%new3__7.pdf: 235x176 pixel, 72dpi, 8.29x6.21 cm, bb=0 0 235 176
\caption{Spatiotemporal dynamics of atomic wave packet in the absence of
modulation (a), and in the presence of modulation (b). The amplitude of
modulation is $\lambda=3$ and $q=85.14$. Other parameters are
$\tilde{V}_{0}=0.36$, $\Delta p=0.1$ and $k^{\hspace{-2.1mm}-}=0.16$.}
\label{fig:Spatiotemp}%
\end{figure}
\begin{figure}[h]
\includegraphics{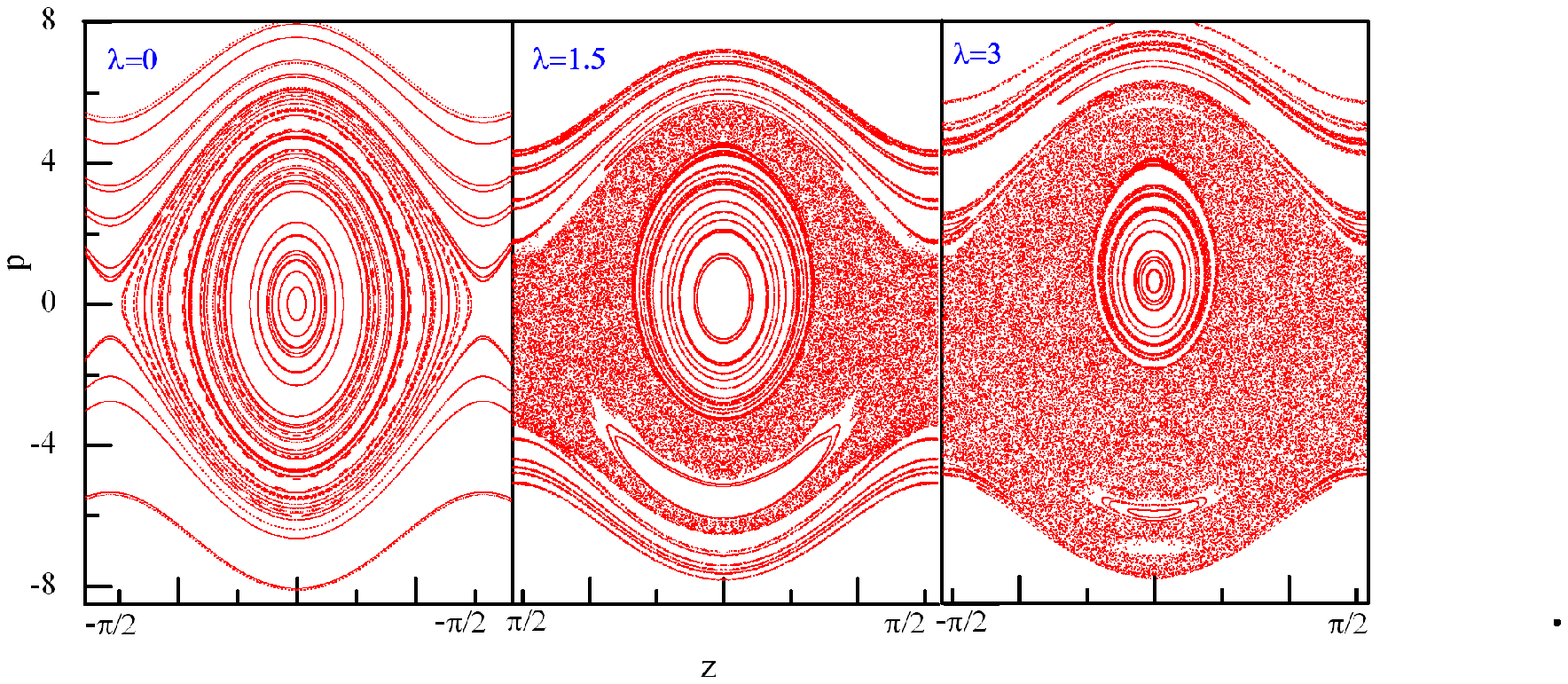}

\caption{Poincar\'e phase space for driven optical lattice for modulation
strengths $\lambda=0,  1.5,  3$ and potential depth $V_{0}=16E_r$}%
\label{fig:FigPhasespace}%
\end{figure}\begin{figure}[h]
\includegraphics{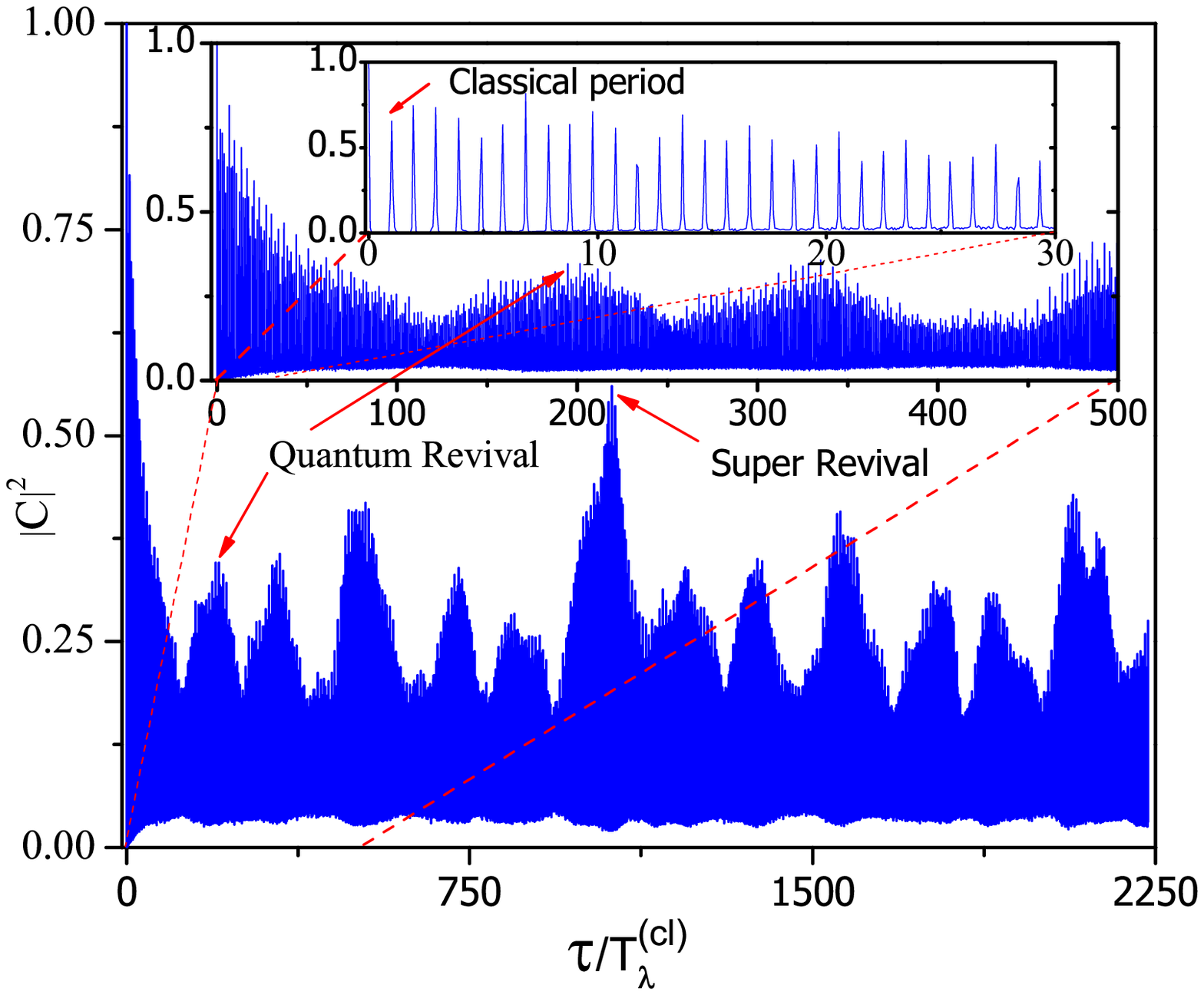}

\caption{Square of auto-correlation function of a Gaussian wave packet is
plotted versus time. Here, the parametric values are $\lambda=1.5$,
$k^{\hspace{-2.1mm}-}=0.5$, $V_{0}=16 E_r$, $\Delta z=\Delta
p=0.5$. The wave packet is initially well localized at the central lattice
well around the second band of undriven lattice. Classical period, revival
time and super revival time are seen and indicated by arrows.}%
\label{fig:Auto15}%
\end{figure}
\end{onecolumn}

\begin{thebibliography}{}
\bibitem{PR1957}P. Bocchieri, A. Loinger, Phys. Rev. 107, 337 (1957).

\bibitem {SaifPR}F. Saif, Phys. Rep. 419, 207 (2005); ibid 425, 369 (2006).
\bibitem{ReinhardtBlumel2001}R. Blumel, W. P. Reinhardt, {\it Chaos in Atomic
         Physics} (Cambridge university press, Cambridge, U.K. 2001).
\bibitem {SaifPRE}F. Saif Phys. Rev. E 62, 6308 (2000).

\bibitem {Saif2005-a}F. Saif, J. Opt B: Quant. Semiclass. Opt. 7, S116 (2005).

\bibitem{SIChuPR2004} S. I. Chu, D. A. Telnov, Phys. Rep. 390, 1 (2004).
\bibitem {RobertPRA2002}R. Luter, L. E. Reichl, Phys. Rev. A 66, 053612 (2002);
G. Hur {\it et al}, Phys. Rev. A 72, 013403 (2005); S. Fishman {\it et al},
Phys. Rev. Lett. 49, 509 (1982); X.
Luo {\it et al}, 2008 Phys. Rev. A 77 053601; ibid 80, 053603 (2009); K. Hijii,
S. Miyashita, Phys. Rev. A 81, 013403 (2010).

\bibitem{VelaPRA2005} Vela-Arevalo Lu V, Fox R. F.,  Phys. Rev. A 71, 063403
 (2005).
\bibitem{HurPRA2005} Hur G. {\it et al}, Phys. Rev. A 72, 013403 (2005).
\bibitem{ZhangPRB2006} Zhang C {\it et al},  Phys. Rev. B 73, 085307 (2006).
\bibitem{VillasPRB2004} Villas-Bˆoas J M {\it et al}, Phys. Rev. B 70, 041302
(2004).
\bibitem{SonPRA2008} S. K. Son, S. I. Chu, Phys. Rev. A 77, 063406 (2008).
\bibitem{Eckardt} A. Eckardt {\it et al}, Phys. Rev. Lett. 95, 260404
(2005).
\bibitem{HolthausPRA2001} M. Holthaus, Phys. Rev. A 64, 011601 (2001).
\bibitem{HensingerPRA2004}  W. K. Hensinger {\it et al}, Phys. Rev. A 70, 013408
(2004).
\bibitem {Saif2002}F. Saif, M. Fortunato, Phys. Rev. A 65, 013401 (2001).

\bibitem {ShahidPLA}S. Iqbal {\it et al}, Phys. Lett. A 356, 231 (2006).
\bibitem{Applications}J. M. Zhang, W. M. Liu, Phys. Rev. A 82, 025602 (2010);
C. Petr {\it et al}, Phys. Rev. E 81, 046219 (2010); M. Heimsoth {\it et al}, 
Phys. Rev. A 82, 023607 (2010); A. Kenfack {\it et al}, Phys. Rev. Lett. 100,
044104 (2008); G. Lu {\it et al}, Phys. Rev. A 83, 013407
(2011); C. Sias {\it et al}, Phys. Rev. Lett. 100, 040404 (2008); C. Weiss, H.
P. Breuer,
Phys. Rev. A 79, 023608 (2009); C. Weiss, N. Teichmann, J. Phys. B 42, 031001
(2009);
M. Roghani {\it et al}, Phys. Rev. Lett. 106, 40502 (2011). 

\bibitem{KurnNature2008}K. W. Murch {\it et al}, Nat. Phys. 4, 561 (2008); K. Zhang {\it et al}, 
Phys. Rev. A 81, 013802 (2010); T. P. Purdy {\it et al}, Phys. Rev. Lett. 105,
133602 (2010);
M. Paternost {\it et al}, Phys. Rev. Lett. 104, 243602 (2010); W. Chen {\it et
al}, Phys. Rev. A 81, 053833 (2010); A. B. Bhattacherjee, Phys. Rev. A 80,
043607
(2009); J. Stettenheim {\it et al}, Nature 466, 86 (2010).
\bibitem{PottingPRA2001}S P\"otting {\it et al}, Phys. Rev. A 64, 023604 (2001).
\bibitem {Bardroff1995}P. J. Bardroff {\it et al}, Phys. Rev.
Lett. 74, 3959 (1995).
\bibitem {Raizen1999}M. G. Raizen, Adv. At. Mol. Opt. Phys. 41, 43 (1999).


\bibitem{Lenz2010} F. Lenz {\it et al}, Phys Rev. E 82, 016206 (2010); ibd New
J. Phys. 11 083035 (2009); W. Acevedo, T. Dittrich, J. Phys. A: Math. Theor.
42, 045102 (2009). 

\bibitem {Lichtenberg}A. J. Lichtenberg, M. A. Liberman, \textit{Regular and
Stochastic Motion}, (Springer, Berlin, 1992).

\bibitem {LindaReichl}L. E. Reichl, \textit{The Transition to Chaos}, 2nd
edition, (Springer, New York, 2004).

\bibitem {E.Ott}E. Ott, \textit{Chaos in dynamical systems} 2nd edition
(Cambridge University Press, Cambridge, 1993).
\bibitem{GoulvenPRA2011} G. Qu´em´ener, John L. Bohn, Phys. Rev. A 83, 012705
(2011).

\bibitem{Oliveira} T.
Gilbert {\it et al}, J. Phys. A: Math. Theor. 44, 065001 (2011); D. F. M.
Oliveira {\it et al},
Physica D 240, 389 (2011).

\bibitem {Flatte}M. E. Flatt\'{e}, M. Holthaus, Ann. Phys. (N. Y.) 245, 113
(1996).

\bibitem {Berry}M. V. Berry, Philos. Tran. R. Soc. Lond. B 287, 237 (1997).


\bibitem {Almeida}A. M. O. de Almeida, J. Phys. Chem. 88, 6139 (1984).

\bibitem {SaifEPJD}F. Saif, Eur. Phys. J. D 39, 87 (2006).

\bibitem {AbramowitzStegun}M. Abramowitz, I. A. Stegun, (eds.) \textit{The
Handbook of Mathematical Functions} (Dover, New York, 1970).

\bibitem {McLachlan}N. W. McLachlan, \textit{Theory and Applications of
Mathieu Functions,} (Oxford university press, London, 1947).

\bibitem {Holthaus1991}H. P. Breuer, M. Holthaus, Ann. Phys. 211, 249 (1991).

\bibitem {Born}M. Born, \textit{Mechanics of Atoms}, (Ungar, New Yark, 1960).
\bibitem {GPBerman1977}G. P. Berman, G. M. Zaslavsky, Phys. Lett. A 61, 295
(1977).

\bibitem {Yukalov2009}Y. I. Yukalov, Laser Phys. 19, 1 (2009).

\bibitem {Moore1994}F. L. Moore, {\it et al}, Phys. Rev. Lett. 73, 2974 (1994).

\bibitem {Holthaus2010}A. Eckardt {\it et al},  Phys. Rev. A 79, 013611 (2009);
S. Arlinghaus, M. Holthaus, Phys. Rev. A 81, 063612 (2010); S. Arlinghaus, M.
Langemeyer, M. Holthaus, in \textit{Dynamical Tunneling - Theory and
Experiment,} (Taylor \& Francis, 2010).

\bibitem {DrytingPRA1993}S. Dryting, G. J. Milburn, Phys. Rev. A 47, 2484
(1993).

\bibitem {AyubJRLR2009}M. Ayub {\it et al}, J. Rus. Laser Res.
30, 205 (2009).

\bibitem {Holthaus1997}K. Drese, M. Holthaus, Chem. Phys. 217, 201 (1997).

\bibitem {Ayub2010}M. Ayub, F. Saif, To be submitted.

\bibitem {ReyPRA2005} A. M. Rey {\it et al}, Phys. Rev. A 72, 033616 (2005).

\bibitem {Ben1996}M. B. Dahan {\it et al}, Phys. Rev. Lett. 76, 4508 (1996).
\end{thebibliography}
\end{document}